# Power law and exponential ejecta size distributions from the dynamic fragmentation of shock-loaded Cu and Sn metals under melt conditions

O. Durand, L. Soulard

CEA, DAM, DIF, F-91297 Arpajon, France

**Abstract:** Large scale molecular dynamics (MD) simulations are performed to study and to model the ejecta production from the dynamic fragmentation of shock-loaded metals under melt conditions. A generic 3D crystal in contact with vacuum containing about $10^8$ atoms and with a sinusoidal free surface roughness is shock loaded so as to undergo a solid-liquid phase change on shock. The reflection of the shock wave at the interface metal/vacuum gives rise to the ejection of 2D jets/sheets of atoms (Richtmyer-Meshkov instabilities in the continuum limit) which develop and break up, forming ejecta (fragments) of different volumes (or mass). The fragmentation process is investigated by analyzing the evolution of the resulting volume distribution of the ejecta as a function of time. Two metals are studied (Cu and Sn) and the amplitude of the roughness is varied. The simulations show that the associated distributions exhibit a generic behavior with the sum of two distinct terms of varying weight, following the expansion rate of the jets: in the small size limit, the distribution obeys a power law dependence with an exponent equal to 1.15 ± 0.08 and in the large size limit, it obeys an exponential form. These two components are interpreted, with the help of additional simple simulations, as the signature of two different basic mechanisms of fragmentation. The power law dependence results from the fragmentation of a 2D network of ligaments arranged following a fractal (scale free) geometry and generated when the sheets of liquid metal expand and tear. The exponential distribution results from a 1D Poisson fragmentation process of the largest ligaments previously generated. Unlike the power law distribution, it is governed by a characteristic length scale which may be provided by energy balance principle.

## I. INTRODUCTION:

The reflection of a shock wave at the interface between the free surface of a shock-loaded metal and its surrounding medium (vacuum or gas) can lead to the ejection of a great amount of small (micron-scale) particles (ejecta). The acquisition of experimental data as well as the development of models and hydrodynamic codes capable to predict the formation of these ejecta and in particular their size or mass distribution is of great importance for the shock compression community. Among all the basic physical parameters which drive the ejection process, experimental and theoretical studies carried on since the 70's have shown that the main are the free surface finish of the sample [1-5], the initial conditions and time loading history of the shock [6-7], and the melting temperature and metallurgical composition of the material [8-9]. From a continuum (hydrodynamic) point of view, ejection is generally postulated as the manifestation of the growth and break-up of Richtmyer-Meshkov instabilities (RMI's) [10-11]. With this approach, the determination of the size or mass distribution of the particles is a very complex task since it requires an accurate knowledge and modeling of the effects of surface tension, viscosity, and dynamic strength of the material in solid and liquid phases [11-12]. Moreover, to cover the whole range of sizes, multiscale and high-resolution hydrodynamic simulations are needed and they are computationally intensive.

In order to give insight of the ejection phenomenon at sub-micron level and sub-nanosecond timescales, atomistic approaches based on MD simulations have demonstrated their great potential [13-16]. Provided that the potential chosen to describe the interactions between the atoms reproduces well the material behavior under shock loading, this approach takes inherently into account all the basic physical mechanisms involved in the process and therefore does not require any modeling approximation as in RMI's based studies. Moreover, it allows following the evolution of the ejection mechanism with time. Of course, MD simulations are limited to small size systems and short time scales but we recently developed a new hybrid method that allows to simulate on reasonable computation times sufficiently large scale systems (>$10^8$ atoms) on long time scales (typically one nanosecond) and that provides ejecta size distributions with a good statistics [17]. This method allowed us also to present some preliminary results of the fragmentation of a shock-loaded crystal of copper with a given amplitude of sinusoidal roughness.

In this paper, we pursue our previous work and we aim to model the dynamic fragmentation of shock loaded metals under melt conditions by analyzing the evolution of the resulting volume (or mass) distribution with time. For this, we investigate the influence of the material and of the amplitude of the roughness on the resulting mass distribution. Two metals are tested: copper (Cu) and tin (Sn) and, for each, the amplitude of the sinusoidal roughness is varied while the wavelength remains constant. The paper is divided as follows. In section II, we describe the generic configuration used for our analysis and we briefly recall the principle of our computation method. In section III, we report the simulation results and in section IV, we present a numerical model. The results are discussed in section V and a conclusion is given in section VI.

## II. DESCRIPTION OF THE SYSTEM AND OF THE COMPUTATION METHOD

Simulations are performed in parallel with typically between 2000 and 4000 cores using STAMP code and TERA-100 petaflopic computer of CEA. A generic configuration consisting in a 3D crystal with a sinusoidal free surface roughness (3 arches) is set in contact with vacuum (see Fig. 1). It is shock loaded so as to undergo a solid-liquid phase change on shock. The time step in all simulations is 1 fs in order to ensure a good conservation of the energy in NVE computation (NVE: microcanonical ensemble in which the total number N of particles, the total volume V, and the total energy E of the system are constant). Periodic conditions are applied in the $O_y$ and $O_z$ directions while the shock propagates in the $O_x$ direction (free condition).

The Cu crystal is face-centered cubic with a lattice constant $a_0$ of 3.54 Å. The [100] direction is parallel to the $O_x$ axis. It is made up 500 × 250 × 250 elementary cells $a_0$ (dimensions: 176 nm × 88 nm × 88 nm) and contains about 125×$10^6$ atoms. We use the embedded atom model (EAM) potential function developed by Sutton and Chen [18] with the parameters of Belonoshko et al. [19]. This potential ensures a good representation of both solid and liquid phases of copper, although the lattice constant is somewhat underestimated with regard to the experiment [20].

For the Sn crystal, we use the EAM potential developed by Sapozhnikov et al. [21]. This potential is optimized to reproduce the properties of shocked Sn that melts from BCC phase according to phase diagram. Therefore, the crystal is BCC with a lattice constant $a_0$ of 3.70097203 Å. The [100] direction is also parallel to the $O_x$ axis. It is made up 300 × 400 × 250 elementary cells $a_0$ (dimensions: 111 nm × 148 nm × 92.5 nm) and contains about $60\times10^6$ atoms.

For each metal, the wavelength $\lambda$ of the roughness is kept constant and is equal to $L_z/3$ ($\lambda$=29.3 nm and $\lambda$=30.8 nm for Cu and Sn respectively); it is considered as almost identical in both cases. The amplitude $h$ of the roughness is varied as also indicated in Fig. 1, and 3 values are tested: 2, 5 and 10 nm. From the point of view of Richtmyer-Meshkov instabilities (RMI's), the corresponding wave number/amplitude product $kh$ ( $k=2\pi/\lambda$ ) for both crystals is ~ 0.42, 1.05 and 2.1 for $h$=2, 5 and 10 nm respectively. The case $kh<1$ corresponds to a linear regime of RMI's and the cases $kh>1$ rather correspond to non-linear regimes of RMI's [12] with extreme conditions of ejection and fragmentation.

The number of ejecta generated during the fragmentation process being less important with $h$=2 nm, 3 additional simulations were performed with this amplitude on both Cu and Sn crystals in order to obtain a good statistics in the particle size distribution. Each additional simulation was launched by varying the seed of the random algorithm which allocates the initial velocities of the atoms around their initial average velocity following a Maxwellian distribution. This procedure ensured to simulate several times the same system with slight different initial conditions and therefore to improve the statistics. The resulting mass distributions were then averaged.

The hybrid method and the analysis tools used in our simulations are described in detail in ref. [17] and we briefly recall their principle. The system is divided in two zones, bulk and surface, which are computed differently during the equilibration stage. The atoms of the bulk zone are brought in a shocked state corresponding to a given density by a quasi-equilibrium MD simulation using the Hugoniostat method [22-23]. The latter offers the main advantage (compared to NVE method) to require a computation time which is virtually independent of the number of atoms and which may be therefore applied to large scale systems. Calculations are performed with compression ratios of the density of 0.65 and 0.7 for the crystals of Cu and Sn respectively. The corresponding thermodynamic properties (particle velocity, shock pressure and temperature) are reported in Tab. 1. These two ratios ensure that both metals are above their respective melting point, in a purely liquid phase. The surface zone is equilibrated simultaneously at 300 K with a near zero pressure. Once the equilibrium is reached in the two zones, the atoms of the bulk zone are set in motion with the particle velocity $u_p$ that corresponds to the shocked state, and the computation goes on for both zones in the microcanonical ensemble NVE.

The fragments/aggregates/ejecta (these terms refer to the same thing in all the following) formed during ejection are analyzed by dividing the system with voxels (volume elementary cells) and by using the Hoshen-Kopelman algorithm [24] to label the atoms that belong to the same fragment. The size distribution of the fragments is studied by using a binning technique [25] which consists in grouping the observed fragments numbers into volume bins of width equal to increasing powers of 2. Large number of few populated volume classes are then clustered together.

The same generic and qualitative behavior was observed for both crystals and, for reasons of clarity, we report below only the detailed study related to the Sn crystal. Quantitative results (in particular particle volume distributions) are nevertheless reported for both crystals and compared.

**III. SIMULATIONS**

### A. The early stages of fragmentation

In Fig. 2, we compare, for the Sn crystal, the qualitative evolution of the system (oblique views) for the 3 roughness amplitudes at the onset of the ejection process, typically 130 ps. The interaction of the planar shock wave with the valleys of the sinusoid gives rise to the ejection of three jets in the $O_x$ direction of propagation of the shock, which develop symmetrically. These jets are in fact sheets of atoms since they are ejected in the ($O_x$, $O_y$) plane due to the extension of the groove in the $O_y$ direction. We note that the elongation of the jets in the direction of propagation of the shock ($O_x$ direction) grows as the amplitude $h$ varies from 2 nm to 10 nm, reflecting a larger asymptotic velocity. This observation is in good agreement with the models and simulations of linear and non-linear Richtmyer-Meshkov instabilities in the continuum (hydrodynamic) limit [12]. As a consequence, the expansion/strain rate of the sheets in the $O_x$ direction grows with $h$ too.

During the expansion of the sheets, holes/cracks, randomly distributed, begin to appear. The density of the holes increases and their time of appearance becomes shorter, around 180-190ps, 100-110 ps and 90-100 ps as $h$ is increased, due to the higher expansion rate. This is observable in top views of Fig. 3a (in all the following, the top view will be relative to the upper atom sheet in the system and is considered as representative of the 3 sheets). In all cases, the holes begin to appear when the mean thickness of the sheets goes down to a value of about 4-5 nm, as indicated in Fig. 3b. This thickness corresponds approximately to twice the "skin thickness" of a sheet. The latter may be roughly defined as the region of a sheet inside which the binding energy of the atoms is influenced by the existence of the surface (surface tension/energy effects). As a sheet expands, its upper and lower skin thicknesses (~ 2 nm) get closer and finally influence each other, making the system highly unstable. In these conditions, the atoms may easily move away from each other and cause the appearance of cracks. We will come back later to this value of 4 nm when we investigate and model the fragmentation of ligaments of liquid metal.

As the time increases, the sheets expand and the initial cracks become pores of void. Considering for instance the case $h$ = 5 nm (the same generic behavior was observed for all amplitudes), we show on different snapshots in Fig. 4 the evolution of the system at 140, 160 and 180 ps. On these snapshots, the pores of void are indicated and colored following their volume. We note that they grow, coalesce and finally percolate in the $O_y$ direction (continuous propagation of the void from top to bottom of the system) around 180 ps. For each initial amplitude, we report in Fig. 5 the log-log volume distribution of the pores at different times before their respective moment of percolation (from 300 ps to 450 ps for $h$=2 nm, from 130 ps to 180 ps for $h$=5 nm and from 80 ps to 130 ps for $h$=10 nm). Whatever the value of $h$ is, the distributions obey a generic steady-state power law form on several decades with almost the same exponent ~ 1.2-1.3. Such a type of distribution is in good agreement with the distributions measured in various processes of growth and coalescence whether of droplets [26] or pores in spallation phenomena [27]. Its existence, before the effective occurrence of the void percolation, suggests, early in the fragmentation mechanism, the establishment of a hierarchical process (growing and coalescence). This point will be very important when we discuss below the origin of the power law in the volume distribution of the aggregates.

**B. Creation of the ejecta**

Once the void has percolated, a complementary network of ligaments of liquid metal between the pores appears. For $h$=2, 5 and 10 nm, Fig. 6 to 8 respectively, show snapshots of the system at the early and late times of the fragmentation process (top and oblique views respectively), together with the corresponding log-log volume distribution of the fragments.

At the early times, just after the moment of void percolation in the atom sheet (~ 460 ps for $h$=2 nm, ~ 190 ps for $h$=5 nm and ~ 150 ps for $h$=10 nm), we note that the ligaments are shorter in length as $h$ is higher, due to the initial density of holes which grows with $h$ (seen in Fig. 3). The volume distributions of these ligaments exhibit two distinct power laws, in the very small size limit (for volumes below approximately 800 Å$^3$), and over, of the form:

$$N(V) = V^{-\alpha} \qquad (1)$$

with $N$ and $V$ the number and the volume of the particles respectively. The first power law, constant with time, is attributed to the noticeable evaporation (atomization) of the liquid around the sheets as already mentioned in ref. [17]. It may cause the coalescence of atoms in very small ejecta but it is considered as insignificant in the ejection process and will be neglected in the following. Over this limit, in the small or intermediate size range, the exponent $\alpha$ is equal to 1.15 ± 0.08. This value of exponent is very close to the one measured previously in the power law distribution of the pores of void.

In the late times (typically above 600 ps), the oblique views of the whole systems show that the ligaments previously created in the 3 sheets have broken up and have reached a final spherical form. The volume distributions of these final aggregates are in a steady state and they also exhibit the same generic behavior. They contain now the sum of 2 distinct terms (neglecting the first part): in the small size limit, they obey the same power law dependence as in the beginning of the process with $\alpha \sim 1.15$ but in the large size limit, they have changed and obey another form of distribution. They are now centered about an average volume $V_0$ that depends on $h$ and they rapidly decrease over this value.

In Fig. 9, we compare the final steady-state distributions (scaled for an easier comparison) corresponding to each amplitude for the Sn and Cu metals respectively and in Fig. 10, we compare, for a given roughness amplitude, the distributions corresponding to each metal. We note that the exponent $\alpha$ is independent of the material (Cu/Sn) and of the initial roughness amplitude since it remains constant and equal to 1.15 ± 0.08. This value is in good agreement with the value found in our previous study on copper [17]. This strongly suggests that the power law arises from a random process. On the contrary, in the last part of the distribution, the average volume $V_0$ depends on the material and on the roughness amplitude: as $h$ increases, it tends towards the decreasing volumes and, for a given initial amplitude, it is larger for the Cu crystal than for the Sn crystal.

**IV. MODEL**

Although it can be observed and experienced in every day life, dynamic (instantaneous) fragmentation is a natural phenomenon, far from equilibrium, which is still difficult to be theoretically well understood and modelled. It applies on all scales (from atom nuclei up to the supernovae) and it has been an intensive subject of research of theoretical and practical interest for many decades. The main challenge is to understand how the underlying basic fragmentation mechanisms involved affect the resulting fragment size (or mass) distribution. The abundance of experimental data and analytical/numerical models in literature shows that the fragment distributions obey mainly two types of basic forms: exponential and power law.

Exponential type distributions are very generally observed in the fragmentation of metals and other more ductile materials [28-31] or in the fragmentation of liquids under homogeneous expansion [32-33]. Their main feature is that they are described by a characteristic/average fragment size, around which they are centred, and over which they rapidly decrease. They are thus tight distributions. In solids, they find their origin in the presence of uncorrelated cracks or failure points that are governed by a Poisson statistics (in 1D or 2D dimensions) [31].

On a striking way, as reported by a large variety of experiments and computer simulations, power law type distributions are very frequently observed in processes that range on all length scales, from the break up of heavy nuclei [34], through the fragmentation of brittle materials such as glass or ceramic (brittle fragmentation) [35-41], and up to the fragmentation of comets on astronomic scales [30]. To be more precise, brittle fragmentation rather leads to distributions that contain a combined form of a power law distribution in the small size range of the fragments and an exponential type distribution (strict exponential or exponential-like including Weibull or gamma functions) in the large size range [40]. A power law differs strongly from an exponential in the sense that it does not exhibit any characteristic length scale: it can span a large number of decades in fragment size and it is thus also called a scale free distribution. It presents an aspect of scale invariance [42], which means that whatever the scale at which the distribution is measured, it remains the same. Instead of a characteristic length scale, a power law is characterized by an exponent, a fractal dimension [42-43], which value, in the case of brittle fragmentation, varies from typically 1 to about 2.5 [44]. Several theoretical approaches have been proposed to correlate such a scattering of values to the existence of different universality classes of fragmentation [37, 41, 45-47]. Some studies suggest that the exponent depends only on the shape of the object in a context of self-organized criticality [48] or that it depends on the dimensionality [49-51] or the macroscopic properties of the system [52] rather than on the material. On the contrary, other studies suggest that the value of the exponent is a measure of the fragility of the material [43] or that it varies logarithmically with the amount of energy deposited in the material [53]. Phase transition mechanisms between "damage" and "fragmentation" phases or in the framework of percolation have been proposed in the case of simple fluids [34] or solids [47, 54-56]. Several authors [36-41, 57-59] had success in interpreting their distributions from a hierarchical crack branching-merging mechanism which allowed them to calculate analytically the exponent's value. Grady [29] recently suggested the possibility that the power law finds its origin in an excess of energy accumulated in the material before the fragmentation occurs and that would be dissipated in cascade on successively finer length scales by a crack branching process, in analogy with turbulence hydrodynamic phenomena, until the initial stored energy is totally dissipated. Therefore, the origin of the power law is far from being closed and it will still be debated and discussed in future papers.

The vast majority of these studies deal with fragment size (or mass) distributions which are not time resolved. To our knowledge, very few studies report time resolved fragment size distributions [50, 58]. Our MD simulations allow such an analysis and we are going to see now that it is very helpful to discriminate and explain the origin of the basic mechanisms of fragmentation.

### A. Origin of the power law in the aggregates volume distributions

More generally, power laws appear not only in physics but also in a wide variety of natural and man-made phenomena like biology, geological sciences, economics, computer sciences or human sciences [42]. In his review, Newman report some of the mechanisms by which power laws can arise [42]. They play a central role in fractal geometries, randomly fluctuating processes, iterative and hierarchical processes, phase transitions and critical phenomena with their associated concept of self-organized criticality. Therefore, the problem we face to explain the origin of the power law in our aggregates volume distributions is to find a physical mechanism that presents an iterative/hierarchical character or that may be characterised by a fractal geometry. By just considering the final (late times) volume distributions of Fig. 6c-7c-8c, as in the majority of studies where the distributions are integrated over time, and by analyzing the corresponding state of the system (Fig. 6b-7b-8b), it is not possible to find such a mechanism. The particles are indeed homogeneously distributed in the cloud and no explicit geometrical pattern can be observed.

In fact, we already noted the existence of a hierarchical process: it is the one originated from the growing and the coalescence of the pores of void, before and at the percolation. However, we are primarily concerned with the complementary form of this void percolation, that we may call the "matter percolation".

Some authors have already studied the percolation problem for the region of space which is the complement of the union of randomly located and overlapping spheres of equal [60-61] and unequal radii [62]. They analysed this problem within the framework of a continuum percolation problem, for which no underlying network is defined *a priori* (which is also our case), and they used a geometrical construction such as the Voronoï tessellation to map the space between the spheres. A Voronoï tessellation is defined with respect to a given set of points in space, referred to here as the centres of the spheres. It consists in partitioning the space into polyhedral regions (in 3D) such that each region is closest to one of the given set of points (centres). They showed that the edges of the Voronoï polyhedra constitute a network of bonds and that the vertices of the polyhedra are the sites of the network. Thus the percolation problem for the complement of the union of these randomly located overlapping spheres is equivalent to a bond percolation problem for the edges of the Voronoï tessellation of the sphere centres.

From these previous studies, we may model an atom sheet as, on one hand, a medium consisting of the union of randomly located and overlapping discs (or spheres) of void, of unequal radii, that have percolated, and, on the other hand, as the complement of this medium, that is the region of space located between the pores, made up of matter (ligaments of liquid metal). Using the above results, we may therefore conclude that the matter forms a network of ligaments of liquid metal that may be considered within the framework of a 2D bond percolation model. This observation comforts the assumption we made in our previous study [17].

We make now the following strong hypothesis: we suggest that this 2D network of links/ligaments presents a fractal character since, when it begins to break up, the distribution of the ligaments (in volume or in length since the links have a certain thickness) obeys a power law distribution, at the early times of fragmentation (see Fig. 6c-7c-8c). As the time increases, the links of the network break up and finally form spherical aggregates under the effects of surface tension. The 2D network disappears but the aggregates keep "in memory" its original fractal skeleton through their volume distribution. The determination of the fractal character of a graph is a complex and costly task which is out of scope of this study. Several algorithms exist [63-64] that will be tested later to confirm our hypothesis.

**B. Origin of the average volume $V_0$**

In order to understand why the final volume distribution of the aggregates is centred about an average value $V_0$, we focus our analysis on the representative case $h$=5 nm, knowing that the same generic behaviour was observed for $h$=2 and 10 nm too. In Fig. 11 to 13, we report the ejecta volume distribution at the beginning (from 200 ps to 220 ps), the middle (from 220 to 400 ps) and the end (from 400 ps to typically 600 ps and above) of the fragmentation process, respectively. As a help, in order to better see the evolution of the distributions, we indicate on these curves with a black line the initial power law that was measured at the very beginning of the process (~ 190 ps). We also report the degree of asphericity of the aggregates calculated from the eigen values of the gyration tensor: for a pure sphere, it is equal to zero and for a cylindrical shape, it tends to 1. In Fig. 11 to 13, we also report some representative top views of the (upper) atom sheet in order to visualize the state of the system at moments at which the distribution is measured.

In Fig. 11, from 200 ps to 220 ps, just after the 2D network has begun to break up, we observe that the part of the power law that changes in the large volume limit mainly concerns the fragments which degree of asphericity is above 0.6 and even close to 1 for the largest one. On the contrary, the smallest volumes for which the power law distribution does not change have a degree of asphericity close to 0, i.e. they are rather spherical in shape. Considering then the 4 largest volumes of the distribution, we note that the number of fragments with the largest volume (cut-off volume) decreases while the distribution for the 3 bins of volume immediately smaller increases. During this phase, the degree of asphericity of the new population of fragments increases (up to 0.95 for the largest one). This means that, once it has been created and during a short iterative process, the original 2D network of liquid metal breaks-up first in fragments which still remain relatively large but which shape tends to be cylindrical. This is directly observable in the top views of Fig. 11 at 200 ps and 220 ps.

After this preliminary phase and up to about 400 ps, Fig. 12 shows that the distribution in the large volume limit changes continuously: it is more and more centred about an average volume $V_0$ and rapidly decreases above, while the degree of asphericity of the aggregates decreases. This means that the (rather) cylindrical ligaments previously generated tend to break up in particles that become more and more spherical. The exception is for the largest ones where the asphericity is still around 0.9. Again, this is clearly observable in top views of Fig. 12 at 260 ps and 300 ps for instance.

On late times, typically above 600 ps, Fig. 13 shows that the asphericity is almost equal to zero for all the populations of aggregates, meaning that they have all become spherical, due to the effects of surface tension. This observation is confirmed by a direct observation of the oblique views of Fig. 13 at 600 ps and 900 ps. The volume distribution is then in a steady state, with two well established and distinct parts in the small and large size limits.

The origin of the cylindrical shape may be explained with Fig. 14 where we note, for time 240 ps, that the fragments with the 4 largest volumes are preferentially orientated following either the $O_x$ direction or the $O_y$ direction (in particular, the 2 largest ones). The orientation following the $O_x$ axis comes from the loading history of the atom sheets: they are indeed mainly stretched following the direction of propagation of the shock. The orientation following $O_y$ reveals on the contrary the fact that no strain is applied in the direction of the roughness groove. So, the largest aggregates correspond to the tip of the atom sheets which has time to get a cylindrical shape due to the surface tension effects and which needs time to break up. This explains why the asphericity of the largest aggregate is still of about 0.9 at 400 ps (see Fig. 12) and why it takes 100 ps more (from 500 ps) to decrease below 0.2 (see Fig. 13).

These observations therefore strongly suggest that, once the initial fractal 2D network has begun to break up, a secondary mechanism of fragmentation takes place. The latter affects the largest fragments which shape may be mainly considered as cylindrical. Therefore, it must obey a 1D random fragmentation/partitioning process following a Poisson statistics as already described by Grady et al [30-31]. In his model based on energy balance, Grady shows that such fragmentation process leads to a particle size distribution that is exponential or exponential like in the case where the number of fragments is large. When the number of breaks within a body is few, then the fragment size distribution depends on the body length and the fragmentation aspects are rather governed by a binomial probability function [30-31]. Since we work with small size systems, our experimental distributions should therefore obey a binomial law. The average mass or volume $V_0$ about which the exponential or binomial distribution is centred should result from the competition between the internal energy stored in the body during its loading and the surface energy required to create a new fracture surface.

Holian and Grady [32] already studied the fragmentation of two- and three-dimensional liquids undergoing a homogeneous adiabatic expansion by MD simulations. They could estimate both internal and surface energies analytically and then deduce an average fragment mass because of the "simplicity" of their interatomic potential (Lennard-Jones pair potential). In our specific case, the EAM potentials used for Sn and Cu are much more complex and a direct analytical expression of both energies is relatively hard to calculate. Therefore, it seemed to us more appropriate to check the validity of our hypothesis by studying directly with additional simple MD simulations the fragmentation process of cylindrical ligaments of metal, brought in the same thermodynamic and mechanical conditions as the atom sheets.

### C. Fragmentation of cylinders

The additional simulations are performed as follows. For each roughness amplitude, we consider that typically the 4 largest volumes in the distribution measured at the first moments of fragmentation may break up in a secondary 1D fragmentation process. The diameter of these cylinders is set at 4 nm since we have seen in paragraph III.A that the fragmentation of the atom sheets begins around this value. The initial temperature of the atoms in the cylinder is equilibrated at the average temperature of the fragments measured at the beginning of the fragmentation process. In Fig. 15, we see that, independently of the fragment volume, it is almost constant, of about 1000K, 1150K and 1250K for $h$=2, 5 and 10 nm respectively; it grows as $h$ is increased, due to the increasing expansion rate of the jets. The initial conditions for the atom velocity are chosen as follows. In Fig. 16, we report for each roughness amplitude, just after the beginning of fragmentation, the difference of velocity between the tip of the atom sheets and the bulk. We note the existence of a gradient with slopes of about 2.8, 6.8 and 9.3×10$^9$ s$^{-1}$ for $h$= 2, 5 and 10 nm respectively. Therefore, as the network begins to break-up and the links/ligaments begin to separate, there exists a velocity gradient between their ends, which is as larger as the ligaments are elongated. In order to reproduce these effects, for each volume of cylinder tested, therefore for each length tested since the diameter is fixed, we impose during a single iteration a velocity pulse at one end of the cylinder (the other one being at rest) determined from the velocity gradient measured in Fig. 16. Note that this gradient corresponds to the maximum of velocity that applies between the ends of the cylinder since it is measured along the $O_x$ axis, i.e along the direction of stretching of the matter. For the cylinders that are oriented differently in the ($O_x$, $O_y$) plane, it takes intermediate values, down to zero for the cylinders orientated following the $O_y$ axis. The latter correspond indeed to the tip of the atom sheets, where no strain is applied. Then we let the cylinder break up; we count the number $n$ of resulting particles and we deduce their average volume (~ initial volume divided by $n$).

The principle of the computations is summarized in Fig. 17, with the example of the roughness amplitude $h$=5 nm. We observe that for the smallest volume tested $V_1$, the cylinder does not break up (the fragmentation results in a single particle) whatever the velocity gradient is. Indeed, its internal energy is not large enough to cause failure. Therefore, the resulting average volume $V_1^{'}$ is equal to $V_1$. This comforts the choice of not considering smaller volume than $V_1$. For the other volumes, as the velocity pulse increases, the fragmentation results in an increasing number of particles (of decreasing average volume). Typically, the third largest volume breaks up the most, in ~ 5 particles of average volume $V_3^{'} = V_3/5$, as the velocity is 400 m/s. The largest volume $V_4$, as no velocity gradient is applied since it corresponds to the tip of the atom sheet, breaks up in 2 particles of average volume $V_4^{'}$.

In Fig. 18, we report these 4 values of average volume in the corresponding final volume distribution (for $h$=5 nm), at 600 ps and above. We note that the average volume $V_4^{'}$ corresponds to the cut-off volume of the final distribution. The 3 other average volumes are of the same order (~ 1.7×10$^5$ Å$^3$). Therefore, the experimental average volume $V_0$ which results from the contribution of all the volumes must be around this value. As mentioned above, from the model of Grady [30-31], the number $n$ of particles being few in all cases, we check if a binomial distribution of the form below fits our experimental volume (or length) distribution:

$$N(V) \propto \left(1 - \frac{V}{V_L}\right)^{n-2}, \qquad (2)$$

with $V_L$ the initial volume of cylinder. In this example ($h$=5 nm) we take the volume for which the fragmentation is maximum, i.e. $V_L$=$V_3$ with $n$=5. We note that the fit is in very good agreement. For the 2 other amplitudes, Fig. 18 shows that the final distributions are also very well fitted by binomial distributions, considering for $h$=10 nm that the cylinder of volume $V_L$ ~ 4.2×10$^5$ Å$^3$ breaks up in $n$=7 particles and, for $h$=2 nm, that the cylinder of volume $V_L$ ~ 3.4×10$^6$ Å$^3$ breaks up in $n$=5 particles. Therefore, our hypothesis is validated.

We must note however that these additional simulations are simplified. In particular, we have considered our cylinders as perfectly homogeneous with no pre-existing defect. In the numerical experiments, the ligaments are rather fragile due to the initial 2D network that applies strains not only following the main axis of the cylinder but also on the sides. Therefore, the number $n$ of fragments that we obtained is just an average number which will not be necessarily exactly the same for fragments with identical volumes but slight different initial conditions.

We understand now why $V_o$ depends on the roughness amplitude and on the material, as noted in paragraph IIIB. Indeed, as the initial amplitude $h$ increases, the strain in the sheets increases and therefore more internal energy is brought to the ligaments to break the bonds between atoms and create new fractures. Only the largest ones may store enough energy to continue to break up. The fragmentation then generates an increasing number of fragments of decreasing average volume $V_o$ as observed in Fig. 9. Moreover, the surface energy of the fragments being directly linked to the properties of the material and therefore to the potential used, a larger average volume $V_o$ is expected for copper than for Sn, which is also observed in Fig. 10.

**V. DISCUSSION**

The approach of modeling the fragmentation process of a shock loaded crystal with surface roughness under melt conditions, using complete MD simulations, proves its great interest. Indeed, it presents two main advantages. It allows to capture the physics of fragmentation since the simulations show that the latter occurs only for atom sheet thicknesses of the order of a few nanometers. Moreover, our approach offers the possibility to follow the evolution of the resulting fragment volume distribution with time and thus to identify two basic mechanisms of fragmentation that occur sequentially. At the limit where the number of particles generated becomes large, the final volume distribution should take the generic form:

$$N(V) = \beta V^{-\alpha} + \gamma(t) e^{-V/V_0}, \qquad (3)$$

with $\alpha = 1.15 \pm 0.08$ and $V_0$ the average volume issued from the fragmentation of the largest aggregates; $\beta$ and $\gamma$ are factors which account for the respective weight of the two terms. The coefficient $\beta$ is set constant since it exists on all stages of the process while $\gamma$ depends on time since it appears only on late times, once the largest cylindrical fragments have been formed ($\gamma(0)=0$).

These results are in good agreement with the combined forms of power law and exponential distributions reported in brittle fragmentation studies, especially the one of Aström et al [40]. However, we differ on several points from brittle fragmentation. The majority of studies discuss about brittle fragmentation of solid objects, at relatively "low" velocity impact, $\leq 200$ m/s, (free fall impact or collision of particles, discs or spheres [40, 44, 46, 55-56]); to our knowledge, very few studies [65] concern brittle fragmentation under high strain-rate loading. In our case, we deal with the fragmentation of metals, in liquid phase, which undergo very high strain rates (several $10^7$ s$^{-1}$). Therefore, we deal with a new domain of fragmentation. Nevertheless, the presence of both power law and exponential distributions comfort the distinction made by Grady et al. [29-30] between equilibrium and non equilibrium fragmentation. Indeed, during the expansion of the atom sheets, the strain is applied so fast that a fracture energy criterion such as the competition between internal and surface energies (used in equilibrium fragmentation) is not relevant. The excess of energy can not be evacuated in a single time and, in a first stage, non equilibrium fragmentation occurs, that gives rise to a power law volume distribution [29-30]. In a second stage, later in time, as the fragments generated are isolated from each other, the excess of energy may be evacuated following an equilibrium fragmentation process (in a 1D geometry), resulting in a volume distribution of exponential type as reported by Grady [31].

In a power law distribution, Turcotte [43] indicates that the fractal dimension is a measure of the fragility of the fragmented material. In our case, we assume that the exponent $\alpha$ is the same for Sn and Cu metals because we are in a liquid phase in both cases; thus the intrinsic fragility of the material must play a lower role in this regime. We may therefore expect that, under melt conditions, the coefficient remains constant for a certain range of metals.

## VI. CONCLUSION

In order to model the dynamic fragmentation of shock loaded metals under melt conditions, we performed large scale MD simulations with a generic configuration of about $10^8$ atoms consisting in a 3D crystal in contact with vacuum and presenting a sinusoidal free surface roughness. The crystal was shock loaded so as to undergo a solid-liquid phase change on shock and the resulting fragmentation process was investigated by analyzing the evolution of the volume distribution of the ejecta as a function of time. Two metals were tested, Cu and Sn, and 3 roughness amplitudes were tested: 2, 5 and 10 nm, with a constant wavelength of about 30 nm. Both linear and non linear regimes of Richtmyer-Meshkov instabilities in the continuum limit were thus covered.

MD simulations prove their great interest since they allow to capture the physics of fragmentation and they offer the possibility to follow the evolution of the fragments volume distribution with time. They allow then to demonstrate that the fragmentation is initiated by stochastic and very local fluctuations of the sheet thickness that cause the germination of pores in the sheet. These pores grow, coalesce and percolate following a hierarchical process. Once this process is achieved, the fragmentation of the atom sheets is driven by two basic and distinct mechanisms that occur sequentially in time. First a 2D network of ligaments of liquid metal appears between the pores of void after they have percolated. We make the strong hypothesis that this network presents a fractal character since, as its links begin to break up, it gives rise to a fragment volume distribution that obeys an almost pure power law form in the early times, with an exponent equal to 1.15 ± 0.08. The power law is thus established first and is the underlying distribution of the final distribution. As the time increases, the largest fragments generated tend to become cylindrical and they have still enough internal energy (stored during the strain loading of the atom sheets) to break up following a secondary mechanism of 1D fragmentation following a Poisson statistics. This mechanism gives rise to an exponential form of volume distribution.

Our MD simulations are at least 3 orders of magnitude smaller than usual scales of ejection phenomena, at hydrodynamic level, and they are restricted for the moment to the atomistic world. Nevertheless, the existence of a power law scaling with a critical exponent that suggests the existence of a fragmentation process *a priori* independent of the scale, and the prediction of the resulting particles size distribution could provide a basis of comparison for further (experimental and/or numerical) studies, in the macroscopic world. Experimental particles size distributions measured with techniques based on holography provides indeed relatively low spatial resolution (about 1.5-2 µm with green laser [66], down to 0.5 µm targeted with UV laser [67]), although they are difficult to implement. Moreover, surface tension and viscosity effects may be implemented in hydrodynamic codes, although the physical values of these quantities are difficult to know in this particular thermodynamic field.

If the geometrical aspects (fractal network and final particles size distribution) between microscopic and macroscopic scales may be compared, the temporal aspects are different, especially through the dependence of the surface tension with the scale. This could lead to different growth rates between both approaches and thus to discrepancies in the final particles size distributions, even if we think that the two fragmentation processes in our MD simulations exist whatever the scale is. This point will have to be addressed in a future work.

In particular, the fragmentation process occurs in the very nonlinear phases of the RMI's which take place at very different times, because the RMI growth rate depends on the initial $kh$ product. Results will have to be compared at similar values of $kh$ product by using in particular a scaled time defined in recent studies [68-69] on RMI's.

Both microscopic and macroscopic simulations are always addressed within the framework of a perfect material that does not contain any defect (joint grain, impurities ...). Only experimental studies will finally be able to validate our results or to suggest if additional physics must be taken into account.

**ACKNOWLEDGEMENTS**

The authors thank Y.-P. Pellegrini for very useful and fruitful discussions on aggregation and fragmentation processes.

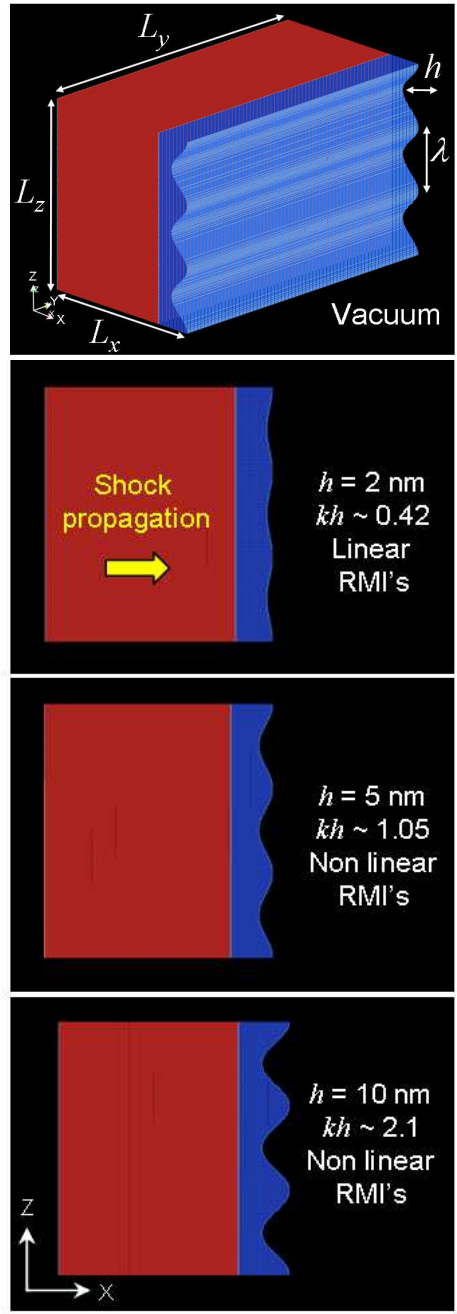

Figure 1: Description of the generic configuration with a sinusoidal free surface roughness. The shock propagates along the $O_x$ axis.

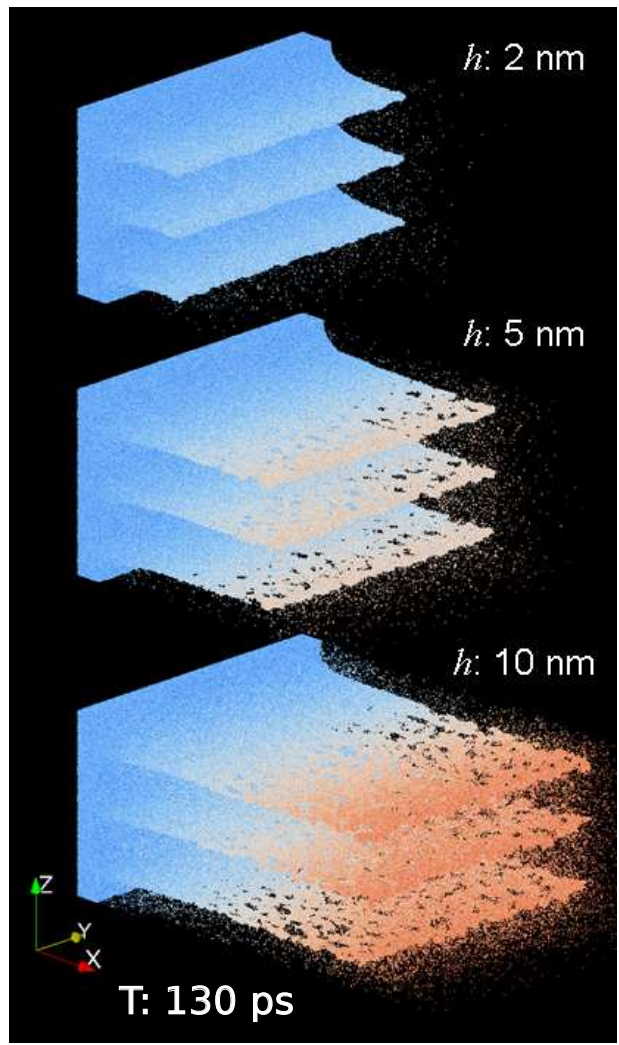

Figure 2: Oblique views of the Sn crystal for the 3 roughness amplitudes at the onset of the ejection process, typically 130 ps.

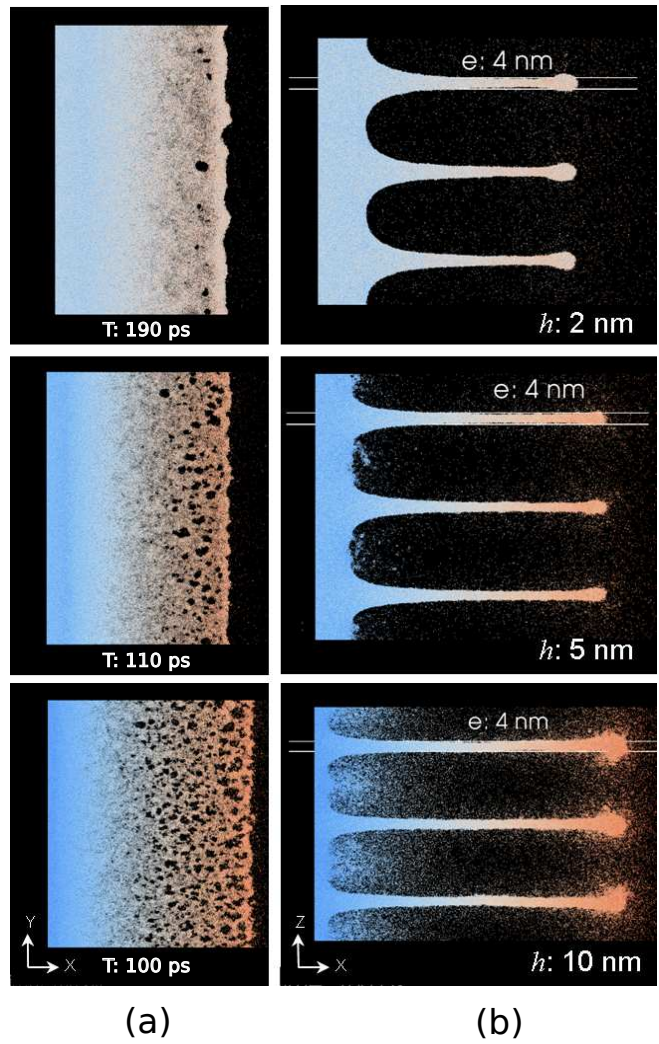

(a)　　　　　　　　　　　　(b)

Figure 3: (a) Top and (b) side views for the 3 roughness amplitudes of the Sn crystal at their respective moment of appearance of holes/cracks in the atom sheets. The sheets begin to break up as their thickness goes down to a value of about 4 nm.

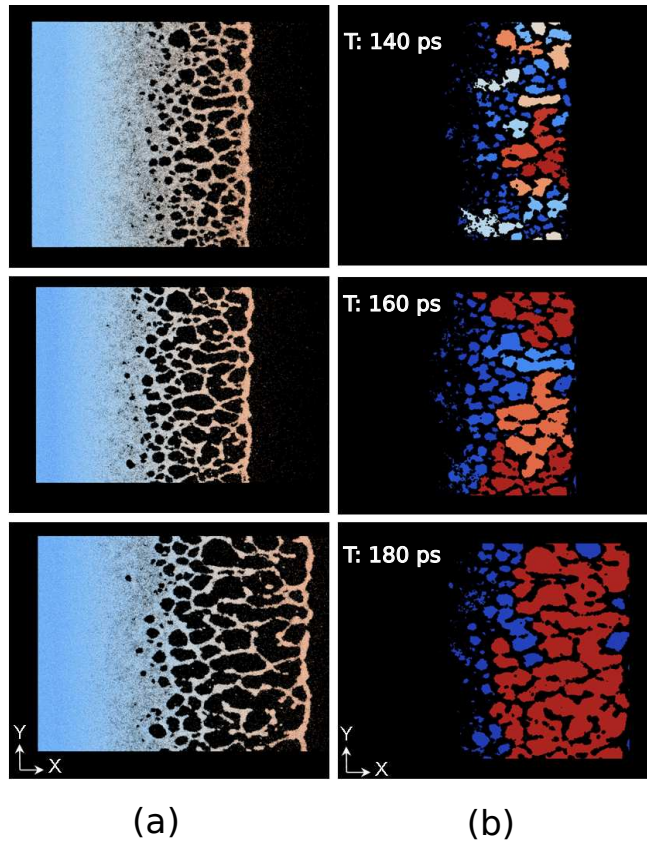

Figure 4: (a) Ligaments of liquid metal and (b) void between the ligaments in the upper atom sheet of the Sn crystal with roughness amplitude $h$=5 nm before (140 ps and 160 ps) and at (180 ps) the percolation threshold. The percolation occurs when the void (indicated in red at 180 ps) can propagate continuously from one end of the sheet to the other in the $O_y$ direction.

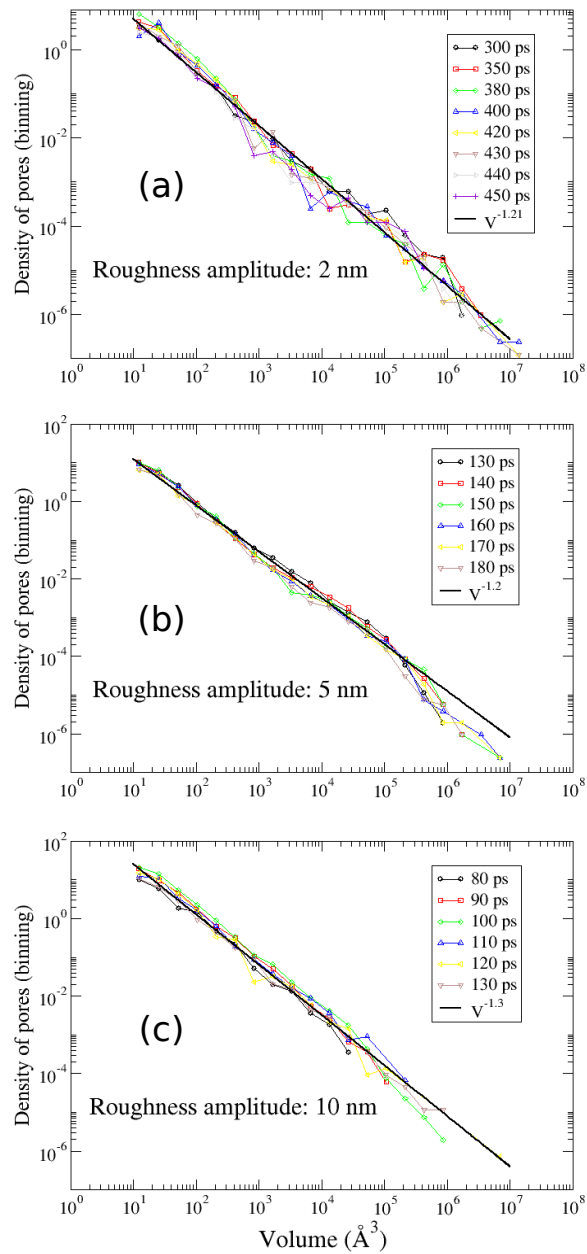

Figure 5: Log-log volume distribution of the pores of void for the Sn crystal with roughness amplitude of (a) 2 nm, (b) 5 nm and (c) 10 nm, at different times before the moment of percolation. All the distributions follow a power law distribution with an exponent ~ 1.2-1.3.

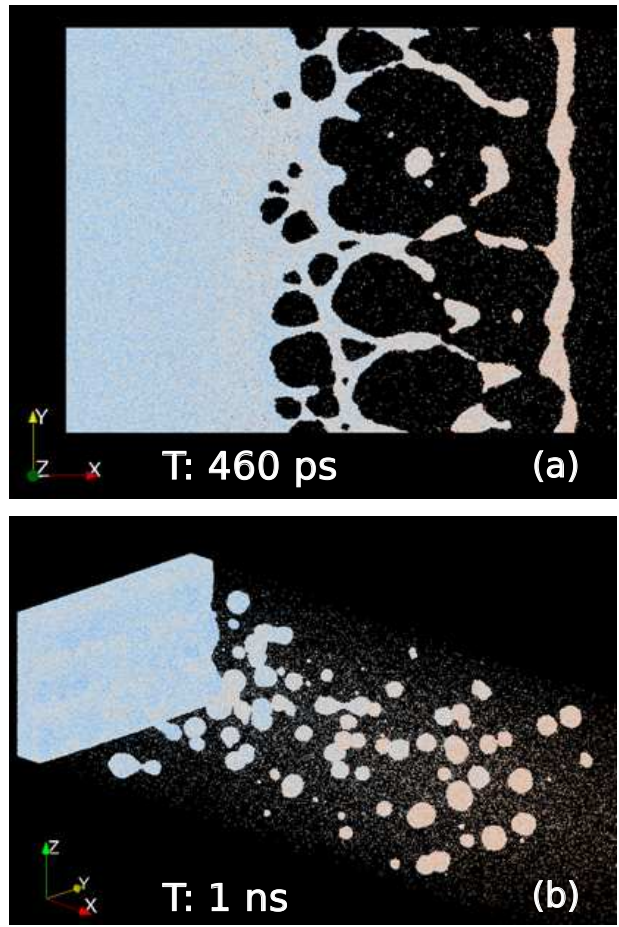
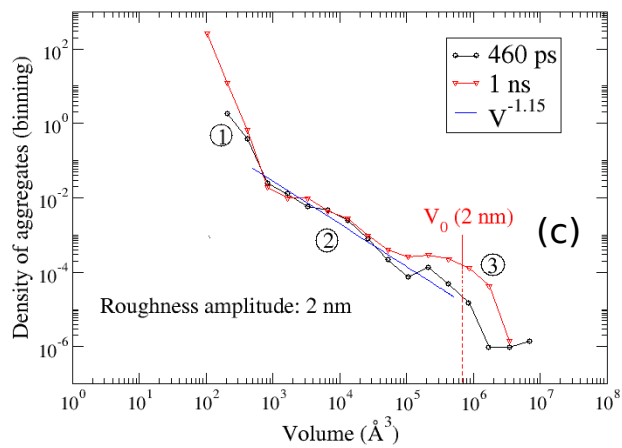

Figure 6: (a) Top view of the upper atom sheet for the Sn crystal with roughness amplitude of 2 nm, just after the void percolation (460 ps), (b) oblique view of the whole system at the late times (1 ns) and (c) corresponding volume distributions of the aggregates.

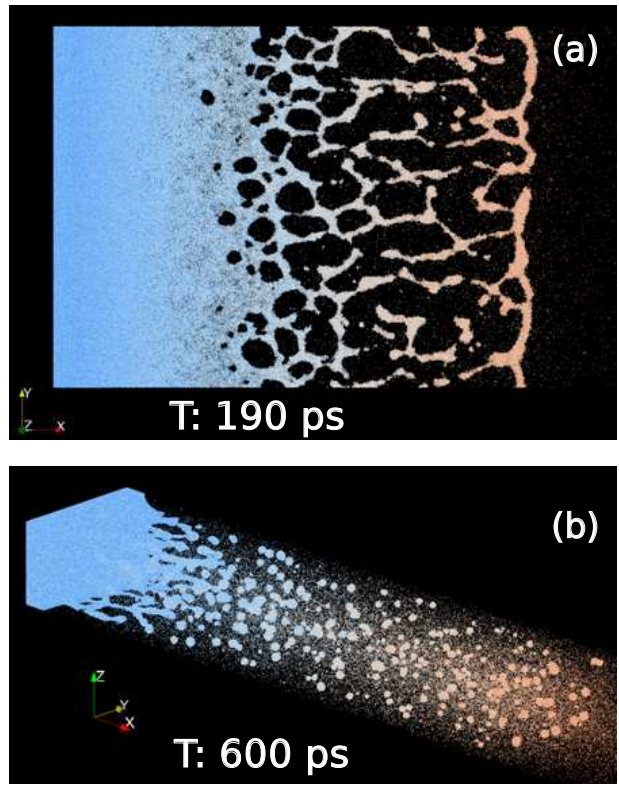
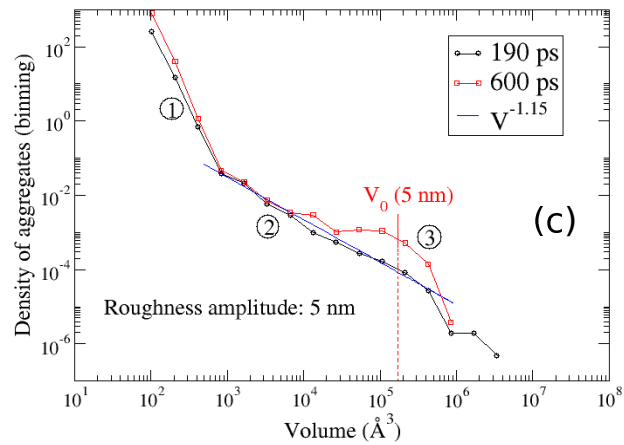

Figure 7: (a) Top view of the upper atom sheet for the Sn crystal with roughness amplitude of 5 nm, just after the void percolation (190 ps), (b) oblique view of the whole system at the late times (600 ps) and (c) corresponding volume distributions of the aggregates.

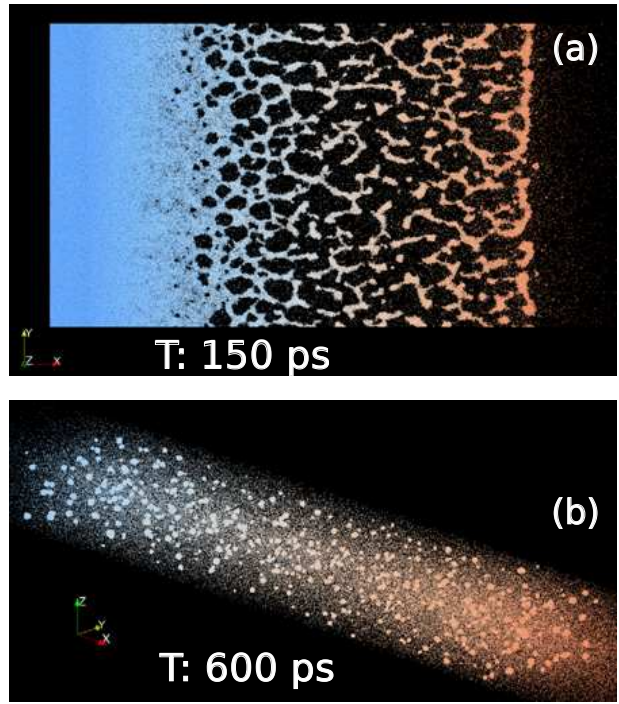
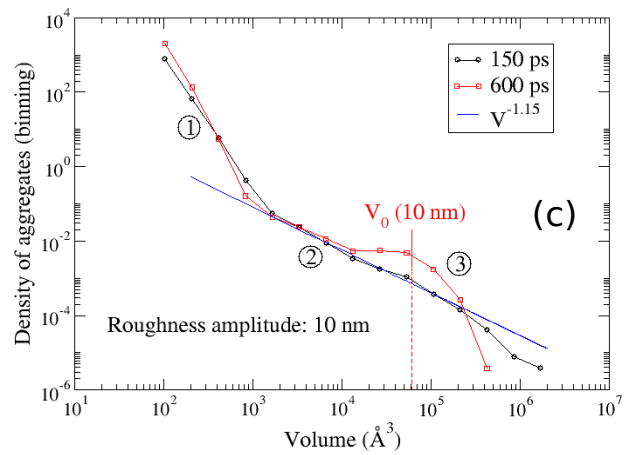

Figure 8: (a) Top view of the upper atom sheet for the Sn crystal with roughness amplitude of 10 nm, just after the void percolation (150 ps), (b) oblique view of the whole system at the late times (600 ps) and (c) corresponding volume distributions of the aggregates.

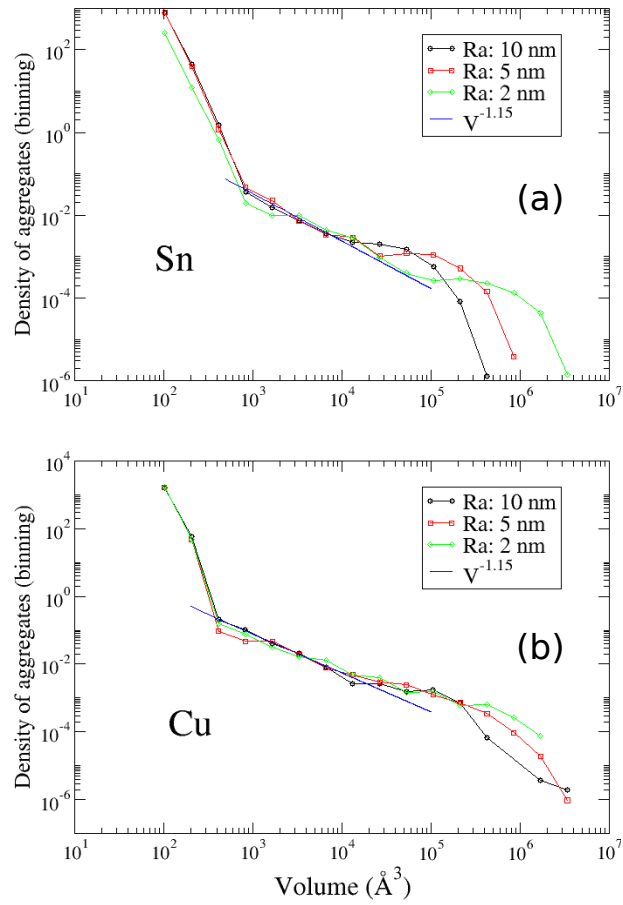

Figure 9: Comparison of the final volume distributions for (a) Sn and (b) Cu crystals as a function of their roughness amplitude.

Final volume distributions of the aggregates

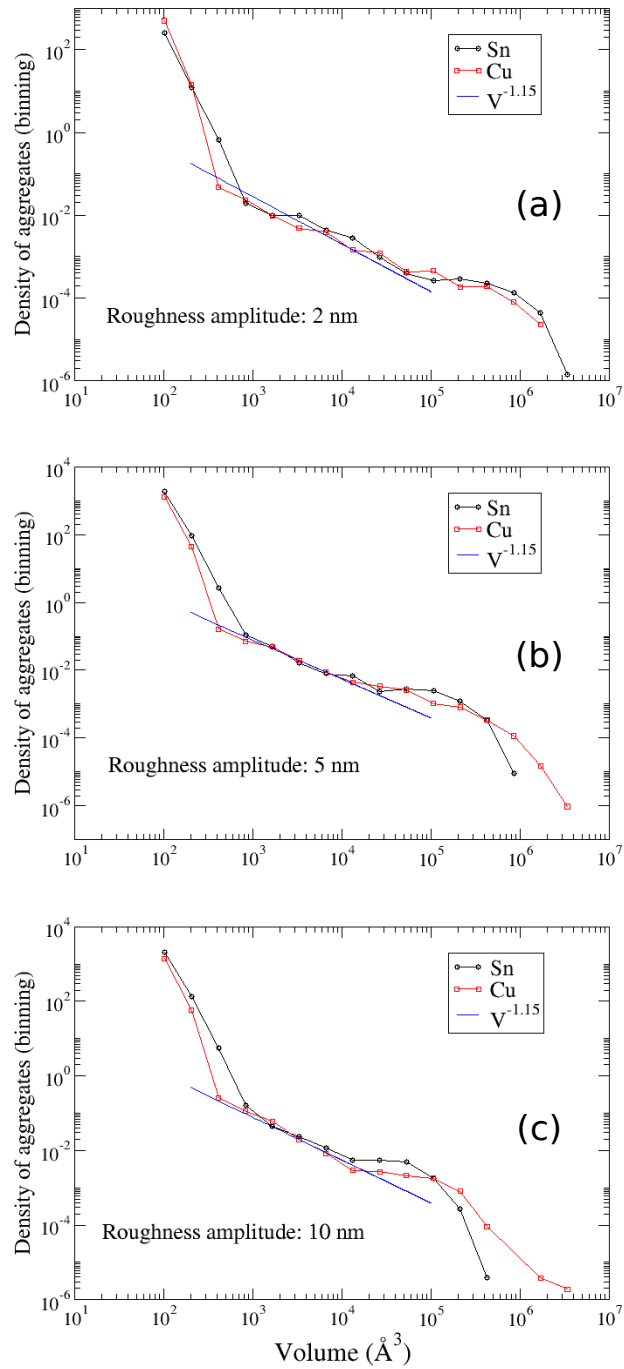

Figure 10: Comparison of the final volume distributions for Sn and Cu crystals with roughness amplitudes of (a) 2 nm, (b) 5 nm, and (c) 10 nm.

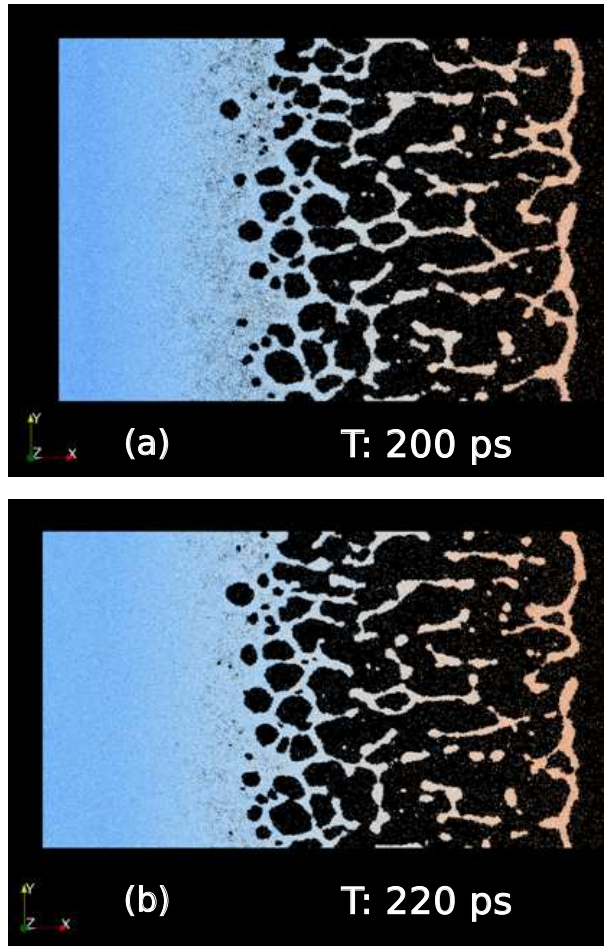

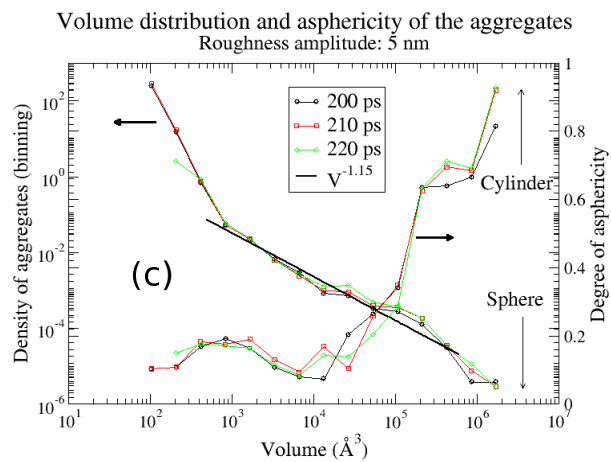

Figure 11: Top views of the (upper) atom sheet for the Sn crystal with roughness amplitude of 5 nm at (a) 200 ps, (b) 220 ps, and (c) volume distribution of the aggregates between 200 and 220 ps. During this phase, just after the beginning of fragmentation of the original 2D network, the ligaments break up in still relatively large aggregates that tend to be cylindrical in shape. The original power law distribution begins to change in the large volume limit.

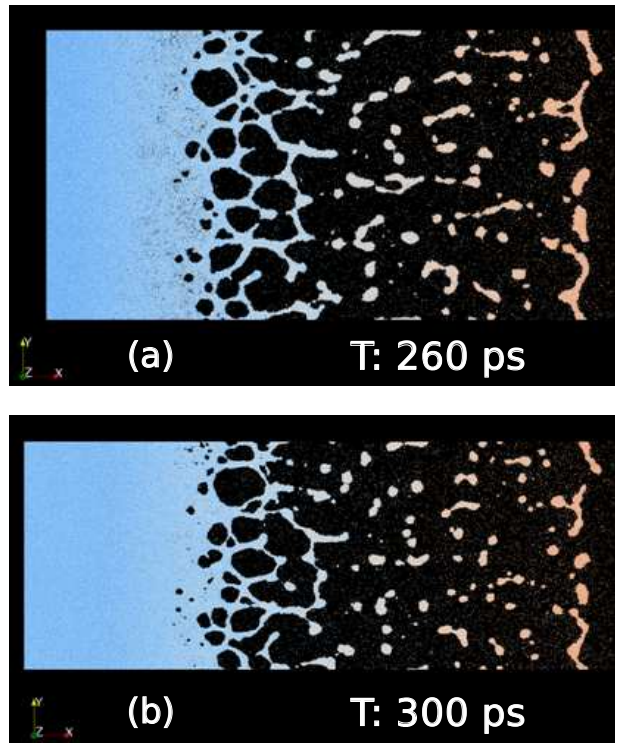

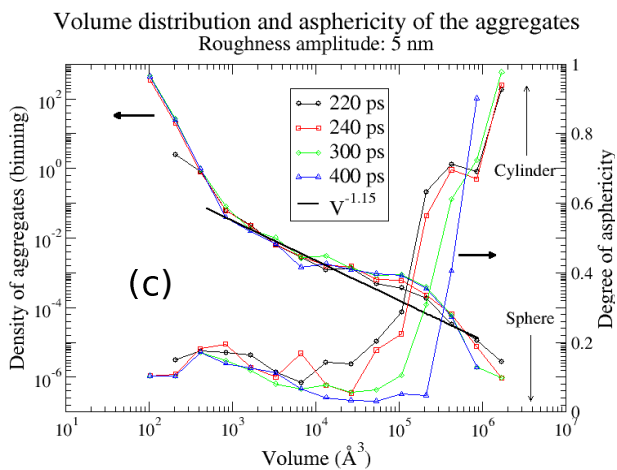

Figure 12: Top views of the (upper) atom sheet for the Sn crystal with roughness amplitude of 5 nm at (a) 260 ps, (b) 300 ps, and (c) volume distribution of the aggregates between 220 and 400 ps. During this phase, the distribution continuously changes in the large volume limit and the aggregates become more and more spherical.

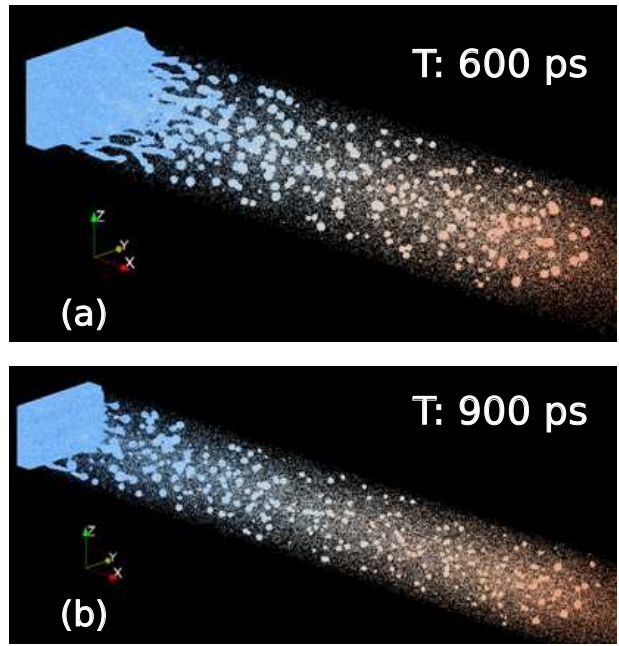
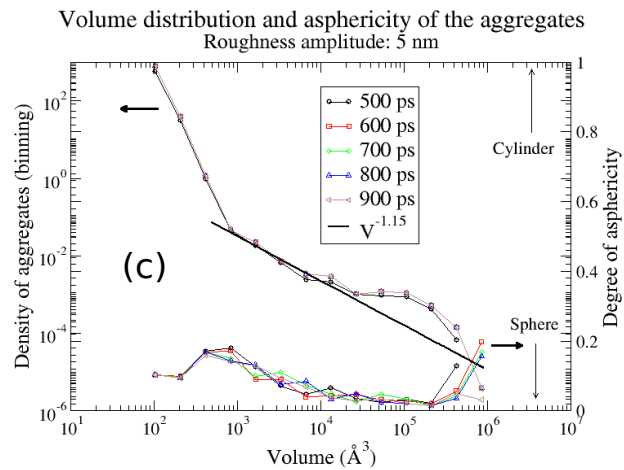

Figure 13: Oblique views of the aggregates ejected from the Sn crystal with roughness amplitude of 5 nm at (a) 600 ps, (b) 900 ps, and (c) volume distributions of the aggregates between 500 ps and 900 ps. On these late times, all the aggregates have become spherical. The distribution exhibits now 2 distinct parts in the small and large volume limits.

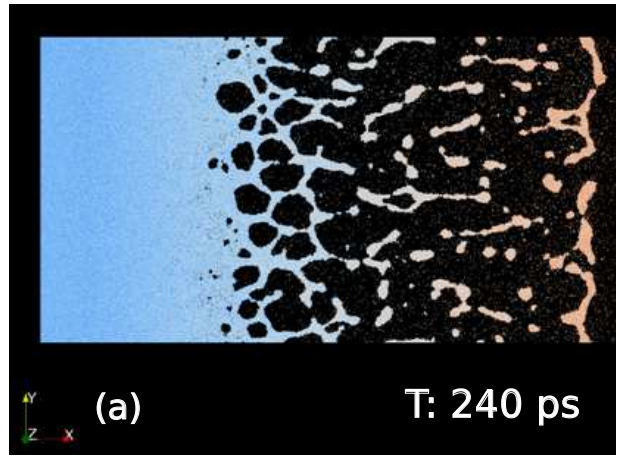

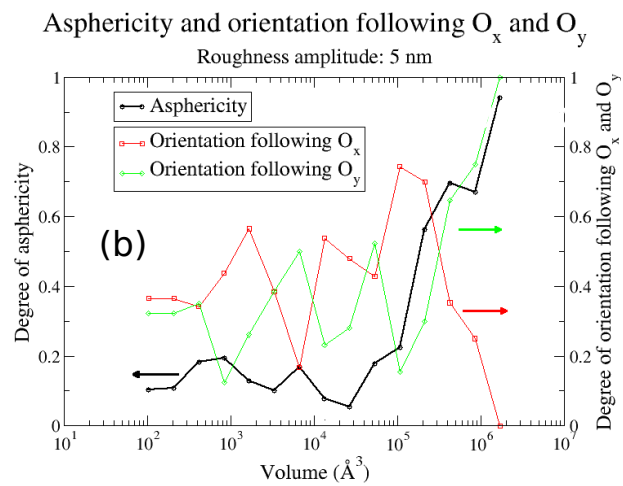

Figure 14: (a) Top view of the atom sheet for the Sn crystal with roughness amplitude of 5 nm at 240 ps and (b) corresponding degree of asphericity of the aggregates with their degree of orientation following the $O_x$ and $O_y$ axes. The largest aggregates have an asphericity ~ 0.95 and are orientated following the $O_y$ axis: they correspond to the tip of the sheet.

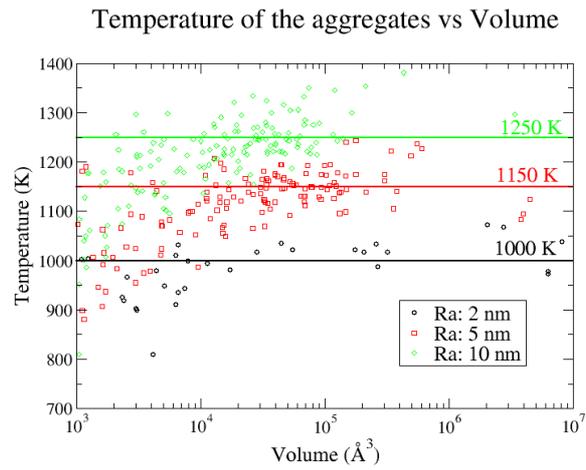

Figure 15: Temperature of the aggregates for the Sn crystal after the onset of fragmentation as a function of the initial roughness amplitude.

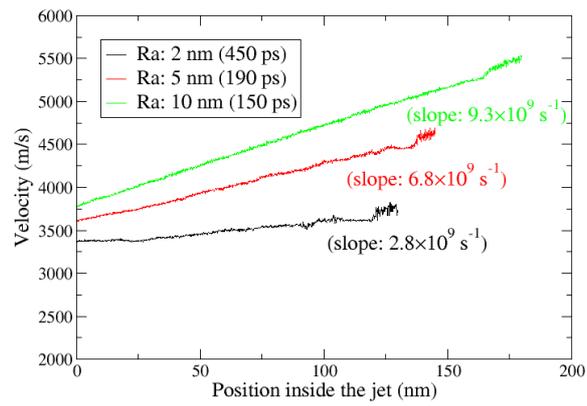

Figure 16: Velocity gradient between the tip of the (upper) atom sheet and the bulk of the Sn crystal for the different roughness amplitudes.

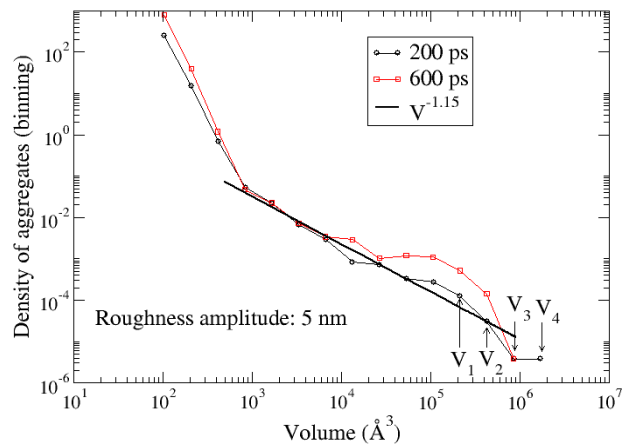
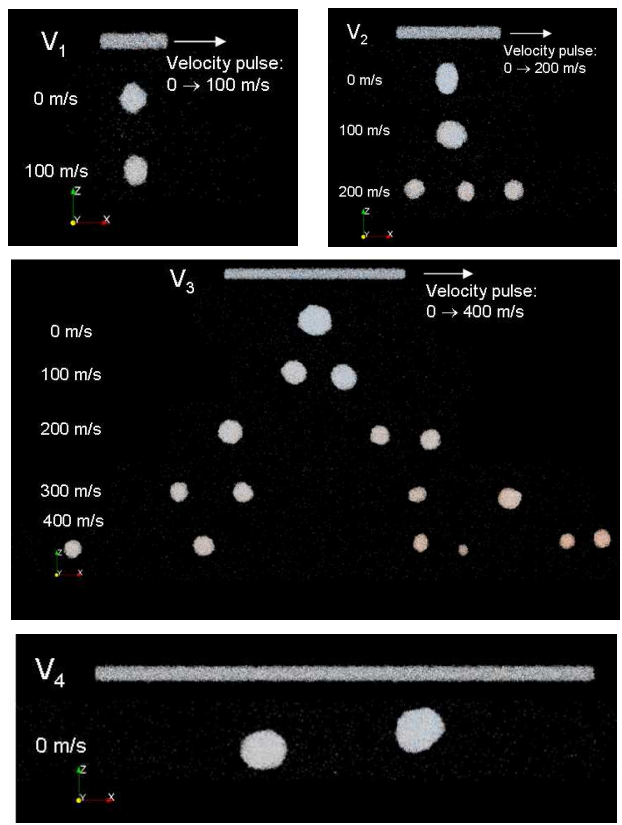

Figure 17: Principle of computation of the fragmentation of cylinders of different volumes with additional MD simulations. The 4 largest volumes of the distribution at the early times of fragmentation are modelled as cylinders with a diameter of 4 nm.

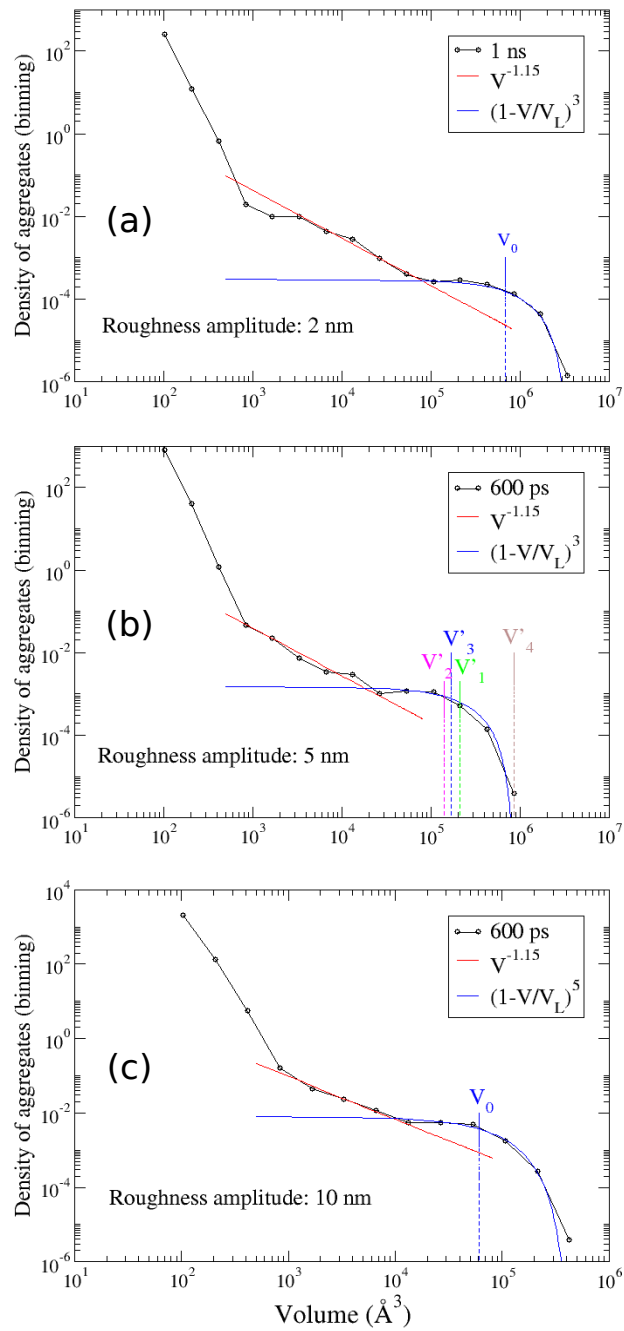

Figure 18: Fit of the final volume distributions with power law and binomial functions for the Sn crystal with roughness amplitudes of (a) 2 nm, (b) 5 nm, and (c) 10 nm.

**Tables**

|  | Copper (Cu) | Tin (Sn) |
|---|---|---|
| Compression ratio: $V_1/V_0$ | 0.65 | 0.70 |
| Velocity particle: $u_p$ (m/s) | ~ 3600 | ~ 1570 |
| Pressure: $P_1$ (GPa) | ~ 360 | ~ 65 |
| Temperature: $T_1$ (K) | ~ 9500 | ~ 2150 |